\documentclass{article}
\usepackage[preprint,nonatbib]{neurips_2019}  

\usepackage[utf8]{inputenc} 
\usepackage[T1]{fontenc}    
\usepackage{booktabs}       
\usepackage{amsfonts}       
\usepackage{nicefrac}       
\usepackage{microtype}      

\usepackage{amsmath, amsfonts, amsthm}
\usepackage{mathtools}
\usepackage{graphicx}
\usepackage{blindtext}
\usepackage[at]{easylist}
\usepackage{todonotes}
\usepackage{algorithm}
\usepackage{hyperref}
\usepackage[noend]{algpseudocode}

\usepackage{import}
\usepackage[shortlabels,inline]{enumitem} 
\usepackage{xspace}
\usepackage{subfigure}
\usepackage{wrapfig}
\usepackage[export]{adjustbox}

\usepackage{xcolor} 
\usepackage{soul} 

\usepackage{ifthen}
\usepackage{xparse}

\newcommand{\any}[1]{\ensuremath{{#1}}\xspace}
\newcommand{\anycmd}[2]{\newcommand{#1}{\any{#2}}}

\newcommand{\ignore}[1]{}

\theoremstyle{plain}    

\theoremstyle{definition}   

\newcommand{\eqn}[1]{\begin{equation}\begin{aligned}#1\end{aligned}\end{equation}}

\graphicspath{{graphics/}}

\anycmd{\N}{N}
\anycmd{\n}{n}

\anycmd{\reals}{\mathbb{R}}
\anycmd{\nnegreals}{\reals_{\geq 0}}

\anycmd{\eps}{\epsilon}

\anycmd{\charfn}{v}
\anycmd{\coopgame}{(\N,\charfn)} 

\renewcommand{\O}{\any{\mathcal{O}}} 
\anycmd{\A}{\mathcal{A}}
\anycmd{\R}{\mathcal{R}}
\anycmd{\T}{\mathcal{T}}
\renewcommand{\S}{\any{\mathcal{S}}} 

\anycmd{\Q}{Q}
\anycmd{\B}{\mathcal{B}} 
\anycmd{\BG}{\mathcal{B}_\mathcal{G}} 

\anycmd{\D}{\mathcal{D}}
\anycmd{\G}{\mathcal{G}}

\title{Decentralized Multi-Agent Actor-Critic with Generative Inference}

\author{%
    Kevin Corder\\ 
    University of Delaware\\
    Newark, DE 19716\\
    \texttt{kcorder@udel.edu}
    \And
    Manuel M. Vindiola\\
    U.S. Army Research Laboratory\\
    Aberdeen, MD 21005\\
    \texttt{manuel.m.vindiola.civ@mail.mil}
    \And
    Keith Decker\\
    University of Delaware\\
    Newark, DE 19716\\
    \texttt{decker@udel.edu}
}

\begin{document}

\maketitle

\begin{abstract}
 
Recent multi-agent actor-critic methods have utilized centralized training with decentralized execution to address the non-stationarity of co-adapting agents. This training paradigm constrains learning to the centralized phase such that only pre-learned policies may be used during the decentralized phase, which performs poorly when agent communications are delayed, noisy, or disrupted. In this work, we propose a new system that can gracefully handle partially-observable information due to communication disruptions during decentralized execution. Our approach augments the multi-agent actor-critic method's centralized training phase with generative modeling so that agents may infer other agents' observations when provided with locally available context. Our method is evaluated on three tasks that require agents to combine local and remote observations communicated by other agents. We evaluate our approach by introducing both partial observability during decentralized execution, and show that decentralized training on inferred observations performs as well or better than existing actor-critic methods.
\end{abstract}


\section{Introduction}
\label{sec:intro}

Reinforcement learning (RL) with function approximation has been used to solve difficult sequential decision making problems in high dimensional state and action spaces, such as game playing \cite{Mnih2013} and robotics \cite{Haarnoja2018}. Many decision making problems are best modeled as a multi-agent system in which agents learn concurrently with other agents. Two naive approaches that use single-agent RL methods in multi-agent problems are independent learning (IL) and joint action learning (JAL), but these approaches perform poorly. IL agents simply treat other agents as part of the stochastic environment.
JAL agents condition on the full joint action and observation spaces for all agents, but these joint spaces grow exponentially with the number of agents and are therefore not scalable. 

Most multi-agent reinforcement learning (MARL) methods use decentralized policies where each agent's policy depends on local observations and actions. Decentralized scenarios usually have partial observations and limited communication. In some problems, the learning must also be decentralized and rely on agent communication \cite{Zhang2018a}. 
Often, a simulator may be available so that agents may learn with extra state information while assuming free, instantaneous communication, e.g., for robotics or autonomous vehicles. 
After centralized learning, the agents execute decentralized policies using only local information. 
Many recent works have adopted this Centralized Training, Decentralized Execution (CTDE) framework; we review several such methods in the following section. 

Agent communication is the main approach to learning in decentralized systems when agents have partial observations. In CTDE methods, the communication during centralized training is implicit: all observations are shared freely and instantly. However, communication networks may be lossy, delayed, or lacking coverage.

We propose to address the limited communication learning gap of CTDE methods in decentralized execution with generative modeling. Specifically, we use a modified context conditional generative adversarial network (CC-GAN) \cite{Denton2016} to infer missing joint observations given partial observations. The task of filling in partial observations by generative inference is similar to the image inpainting problem for a missing patch of pixels: with an arbitrary number of missing observations, we would like to infer the most likely observation of the other agents. 

We extend the popular MADDPG method \cite{Lowe2017} as it appears most amenable to full decentralization. MADDPG agents require both (1) the policies of other agents and (2) other agents' observations as input to the policies and critics. As other agents' approximate policies can be learned, the agents only need to learn a model for the other agents' observations. The generative model will learn this joint distribution of agent observations by training on random combinations of missing agent communications during centralized training. During the decentralized portion, the agents may sample from this model to continue learning under partial observability. 

Our contributions are as follows. We review the recent trend of CTDE MARL literature and identify that they are ill-suited to learn in the decentralized execution phase without explicit communication. We show how a context conditional generative model can address this problem for a popular CTDE method and provide a modified GAN that can learn the joint observation distribution. We experimentally evaluate our approach on three continuous multi-agent tasks. To the best of our knowledge, this is the first work to use generative models to overcome multi-agent partial observability or address decentralized learning in CTDE methods.

\section{Related work}
\label{sec:related}

\subsection{Centralized Training with Decentralized Execution}

Many recent multi-agent actor-critic methods utilize centralized training with decentralized execution. This training procedure lessens the non-stationarity of co-adapting agents by providing additional information in centralized training. The majority of these methods: (1) solve only cooperative tasks by using a shared reward for all agents, (2) use a centralized critic function that conditions on all agents' observations and actions, and (3) use policy or critic networks that include recurrent components, such as an LSTM \cite{Schmidhuber1997}. Recurrent networks have been shown to be effective at learning policies in partially observable environments \cite{Hausknecht2015}. 

Gupta et al. solve cooperative, partially observable tasks with recurrent policies and curriculum learning \cite{Gupta2017}. 
They compare several versions of the CTDE methods, including using Q-learning vs. actor-critic, centralized vs. decentralized policies with parameter sharing, and feed-forward vs. recurrent policies. 
Foerster et al. use a centralized critic for all agents with decentralized recurrent policies for cooperative tasks \cite{Foerster2018a}. 
In addition, they use counterfactual baselines: difference rewards comparing agents' actions to a default action. 

Instead of learning a single centralized critic for all agents, Lowe et al. introduced multi-agent DDPG (MADDPG) which has a centralized critic for each agent and may be used in cooperative or competitive tasks \cite{Lowe2017}. While recurrent networks may be used with MADDPG, only feed-forward networks were tested.
Rashid et al. also uses centralized critics for each agent, but includes a centralized mixing network to combine each agent's critic function \cite{Rashid2018}. They also use recurrent polices and may be used in cooperative tasks. 
Foerster et al. learns communication protocols over a limited-bandwidth communication channel \cite{Foerster2016}. They propose two approaches that use recurrent policies in cooperative settings via parameter sharing or sending gradients over the communication channel. 

We chose to extend MADDPG because it appears the most amenable to decentralization: each agent $i$ has its own critic function $Q_i$ (with no mixing network), policy $\pi_i$, approximate policies of other agents $\mu_i^j$, and reward function to allow for both cooperative and competitive tasks. In addition, the policies are deterministic which allows for continuous state and action spaces.

\subsection{Decentralized Learning}

Traditional decentralized MARL approaches rely on persistent reliable communication so that agents may share local observations and jointly choose actions in an uncertain environment. When states are represented in a factored form, agents may solve a distributed constraint optimization problem over the network to choose a good joint action for all agents \cite{Zhang2013}. In pure MARL approaches, agents choose actions while sharing information over the communication channel. In some systems agents share local rewards but the state is fully observable \cite{Zhang2018a}, and others use a communication-based consensus protocol to agree on a global state from local observations before choosing joint actions \cite{Zhang2018}. In contrast, our approach aims to allow learning despite disruptions in communication. When communication is unavailable, we infer the missing data given the available local observations of neighboring agents.

\section{Background}
\label{sec:background}

\subsection{Reinforcement Learning}
\label{ssec:rl}

Formally, each task in decentralized MARL is represented by a discrete-time partially observable Markov Game \cite{Littman1994}, a multi-agent extension of the Markov decision process \cite{sutton2018reinforcementbook}. A Markov Game is a tuple $\langle \S, \A, \O, \R, \T, \n, \gamma \rangle$ where a set of \n agents choose actions based on local observations to maximize their own expected cumulative reward. At each time step $t$, the environment has a true state $s \in \S$ and each agent $i$ simultaneously chooses an action $a_i$ from their individual set of available actions $\A_i \in \A$. The environment stochastically transitions to a new state $s'$ given by the state transition function $\T$, and each agent then receives a reward $r_i$ according to its own reward function $\R_i: \S \times \A \mapsto \reals$. The discount factor $\gamma$ is used for calculating expected return $ R_i = \sum_{t=0}^{T} \gamma^t r_i^t $ for time horizon $T$. Each agent receives a private observation $o_i \in \O_i$ correlated with the state $s$. Agents choose actions using a stochastic policy $\pi_i: \O_i \times \A_i \mapsto [0,1]$, where has parameters $\theta_i$. 

The three main approaches to RL are action-value methods, policy gradient methods, and the actor-critic hybrid approach \cite{sutton2018reinforcementbook}. Q-learning estimates the action-value function $Q^\pi(s,a)$: the future discounted reward when taking action $a$ from state $s$ while following policy $\pi$. Deep Q-Networks (DQN) used Q-learning with neural network function approximation to play Atari games from pixels \cite{Mnih2013}. 
DQN also introduced two stability improvements: target \Q functions that are updated less frequently, and an experience replay buffer that stores environment transitions $(s,a,r,s')$ for decorrelated batch updates.

Instead of learning a value function, policy gradient methods learn a parameterized policy directly \cite{sutton2018reinforcementbook}. 
This approach is often more efficient but tends to have high variance. To reduce variance, actor-critic algorithms combine an action-value function $Q$ along with the parameterized policy $\pi$ to guide policy updates. 

Deterministic policy gradient (DPG) methods learn a policy $\pi: \S \mapsto \A$ that returns a single action \cite{Silver2014}. Deep DPG (DDPG) is an off-policy model-free actor-critic algorithm that combines the DQN value function with a deterministic policy \cite{Lillicrap2016a}. Like DQN, DDPG uses experience replay and target networks for both value and policy networks. Random noise is added to the policy's output for better exploration. From here on, all policies $\pi$ are assumed deterministic.

\subsection{Generative Adversarial Networks}
\label{ssec:gans} 

Generative models learn a data distribution and can generate new samples similar to the learned distribution. The most popular class of models is the generative adversarial network (GAN) \cite{Goodfellow2014}. A GAN is composed of two neural networks with opposing goals: a generator network \G that receives noise as input and produces samples similar to the data distribution, and a discriminator network \D that tries to determine real data points from those sampled from \G. While GANs have largely been applied to image generation, they should be able to learn any joint data distribution. 

Wasserstein GANs (WGANs) are a variant of GANs that have been shown to have more reliable convergence and less mode collapse \cite{Arjovsky2017}. WGAN uses a critic rather than a discriminator (outputs are not probabilities), trains using a simple loss metric that approximates the Wasserstein distance when the network enforces a 1-Lipshitz constraint, and allows pre-training the critic to optimality. To avoid confusing the WGAN critic with the critic \Q, we will continue to refer to it as the discriminator \D as this distinction makes no difference in our work. 

Our work uses the context-conditional generative model, where the model takes a partial input and must generate a complete data sample. The closest computer vision task to our problem is image inpainting, where a patch of pixels from an image is removed and the model must fill the missing patch based on its learned model of the pixels' joint distribution. The CC-GAN objective function is given by
$
    \min_{\G} \max_{\D}\ \mathbb{E}_{\mathbf{x} \sim \mathcal{X}} [\log \D(\mathbf{x})] +\; 
    \mathbb{E}_{\mathbf{x} \sim \mathcal{X}, \mathbf{m} \sim \mathcal{M}}
    \left[\log \left(1 - \D \left(\mathbf{x}_{\mathbf{I}}\right)\right)\right] 
$
where $\mathbf{m}$ denotes a binary mask used to drop a patch from  image $\mathbf{x}$, and $\mathbf{x}_{\mathbf{I}}=(1-\mathbf{m}) \odot \mathbf{x}_{\mathbf{G}}+\mathbf{m} \odot \mathbf{x}$ is the combined inpainted image where $\odot$ is element-wise multiplication.

Other generative models for image inpainting include autoregressive models and context encoders, but they are not suitable for our approach. Autoregressive models, such as PixelRNN \cite{VanOord2016}, require a pre-specified ordering over the pixels and thus will not work for arbitrary missing data. Context encoders use a variational autoencoder coupled with adversarial loss \cite{Pathak2016}, but results tend to be less accurate compared to CC-GANs.

\section{Decentralized Fine-Tuning}
\label{sec:finetuning}

\begin{figure}
    \centering
    \subfigure[]{\label{fig:GAN_update_diagram}\includegraphics[width=0.5\textwidth]{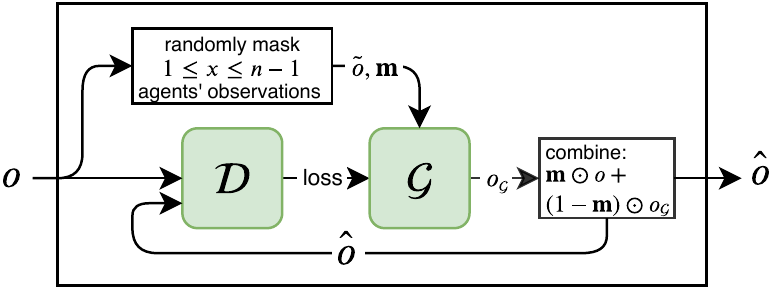}}
    \
    \subfigure[]{\label{fig:my_diagram}\includegraphics[width=0.45\textwidth]{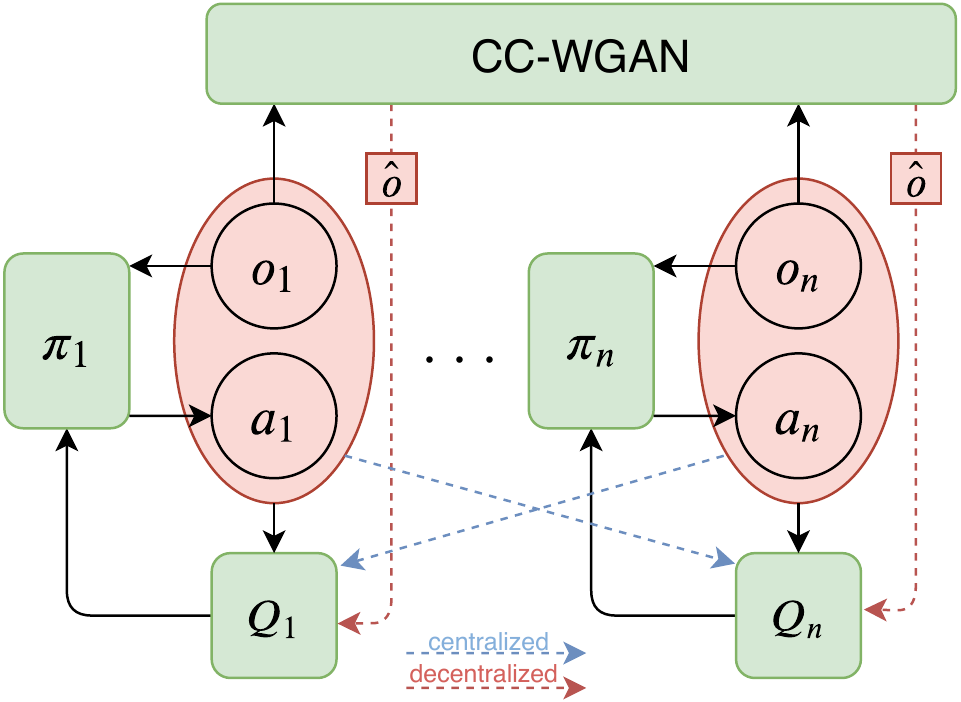}}
    \caption{
    CC-WGAN and MADDPG with inference update diagrams. 
    (a) CC-WGAN agent-based communication training procedure using random masking.
    (b) MADDPG with CC-WGAN learning diagram. Centralized: actor-critic learning updates are identical to MADDPG and CC-WGAN collects joint observations. 
    Decentralized: agents sample from CC-WGAN to fill partial observations. Note that this diagram excludes the approximate policies $\mu$. 
    }
    \label{fig:my_label}
\end{figure}

\begin{figure}[ht]
\centering
\subfigure[]{\label{sfig:obs_vectors}\includegraphics[width=0.45\textwidth]{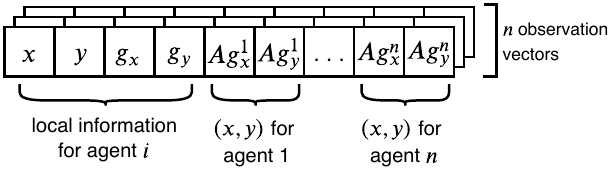}}
\ 
\subfigure[]{\label{sfig:obs_masks}\includegraphics[width=0.45\textwidth]{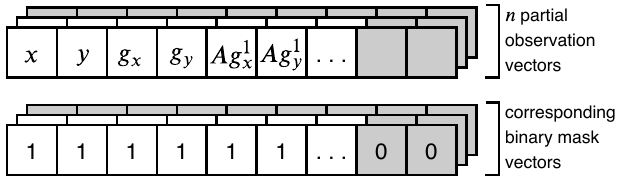}}
\caption{
Random masking input and output used in training the CC-WGAN for an agent-distance partial observability criterion.
Fig. \ref{sfig:obs_vectors} depicts \n agents' observation vectors fed to the masking function. Fig. \ref{sfig:obs_masks} shows the masking function output: a set of \n partial observations and \n binary masks for agent $i$. Masked values in observations are filled with random normal values $z \sim \mathcal{N}(0,1)$. 
}
\label{fig:obs_masking}
\end{figure}

\subsection{Inferring Observations with CC-WGAN}

We approach the problem of inferring missing information from partial observations as a generative sampling problem, similar to the task of image inpainting. We use a modified CC-GAN as our generative model \cite{Denton2016}. Specifically, we train a WGAN with gradient penalty constraints \cite{Arjovsky2017,Gulrajani2017} with the CC-GAN random data masking training procedure. We refer to our modified model as the context-conditional WGAN (CC-WGAN). In our experiments, the CC-WGAN was more reliable than regular CC-GAN for low-dimensional data. Unlike the standard image generation task for GANs, we have no training data. We store joint observations $o = \langle o_1, \ldots, o_\n \rangle$ in a replay buffer \BG just as MADDPG to stabilize learning when batch training. 

Fig. \ref{fig:GAN_update_diagram} shows the update procedure for training the CC-WGAN. When the model updates, it randomly samples joint observations $o \sim \BG$ from the joint observation replay buffer \BG to randomly mask $1 \leq x \leq \n - 1$ agent observations. During centralized training, all joint observations are available. If the CC-WGAN is updated in the decentralized phase, inferred observations are mixed into the updates. This is a form of semi-supervised learning because the model updates on its own predictions \cite{Goodfellow2016book}. 

For each joint observations $o$, we randomly mask combinations of missing agents from $o$ with a binary mask $\mathbf{m}$. 
Masked elements in $o$ are replaced with random normal noise $z \sim \mathcal{N}(0,1)$ to get the partial observation $\tilde{o}$. 
The masking procedure requires some knowledge of the conditions when observations will become partial, e.g., inter-agent distance greater than communications allow. Fig. \ref{fig:obs_masking} shows the masking procedure for distance-based partial observability based on $(x,y)$ coordinates. In the diagram, \G takes joint partial observation $\tilde{o}$ and binary mask vector $\mathbf{m}$, and produces the generated output $o_\G = \G(\tilde{o}, \mathbf{m})$. We then replace the masked portion in $o$ with that portion of the generated output $o_\G$ to get a combined observation $\hat{o} = \mathbf{m} \odot o + (1-\mathbf{m}) \odot o_\G $.

Where CC-GAN passes only the inpainted patch to the discriminator, we instead pass the combined observation to the discriminator because any number of agents may be missing from $\hat{o}$. 
\D is trained to distinguish batches of size $m$ of real joint observations $o$ and inferred observations $\hat{o}$ by minimizing the empirical Wasserstein distance: 
\eqn{ 
    \frac{1}{m} \sum_{i=1}^{m} \left [\D \left(o^{(i)}\right) - \D \left( \hat{o}^{(i)} \right) \right]
}
where $\hat{o} = \mathbf{m} \odot o + (1-\mathbf{m}) \odot \G(\tilde{o}, \mathbf{m}) $. Similarly, \G is updated by maximizing: 
\eqn{
    \frac{1}{m} \sum_{i=1}^{m} \D \left( \hat{o}^{(i)} \right)
}

\subsection{MADDPG with Inferred Observations}

As stated before, we augment MADDPG \cite{Lowe2017} with the CC-WGAN because it appears the most flexible CTDE method for decentralization. Each MADDPG agent $i$ learns a deterministic policy $\pi_i$, a centralized critic $Q_i$, and a set of approximate policies $\mu_i^j$ for each other agent $j$. 

Fig. \ref{fig:my_diagram} shows the MADDPG method updates along with the CC-WGAN for both centralized and decentralized phases. 
During centralized training, the critics $Q_i$ and policies $\pi_i$ are updated exactly as MADDPG. In addition, the CC-WGAN is collecting joint observations in its replay buffer and updating as described above. 

In the decentralized phase, local observations may be missing information about other agents. 
At each time step each agent $i$ receives a partial observation $\tilde{o_i}$ which consists of the agent's local information and possibly information about other agents. When updating the centralized critics, if an agent $i$ has information about agent $j$ in its local partial observation $\tilde{o_i}$, then it can also see agent $j$'s partial observation $\tilde{o_j}$. This is because we assume agents within range are ``communicating'' all local information. 
Just as in training, the joint partial observation $\tilde{o}$ is passed to the generator to get $o_\G$ and combined via binary mask $\mathbf{m}$ with $\tilde{o}$ to get the inferred observation $\hat{o}$. 

Following the derivation in \cite{Lowe2017}, the deterministic policy loss with inferred observations is: 
\eqn{ 
    \nabla_{\theta_{i}} J\left(\pi_i\right) = 
    \underset{\hat{o}, a \sim \B}{\mathbb{E}}
    \left[\nabla_{\theta_{i}} \pi_{i}\left(a_{i} | \hat{o_i}\right) \nabla_{a_{i}} Q_{i}^{\pi}\left(\hat{o}, a \right) \right ], 
} 
where $a = a_1, \ldots, a_n$ are taken from approximate policies such that $\mu_i^j(\hat{o_j}) = a_j$. The approximate policies are updated with: 
\eqn{
    \mathcal{L}(\theta_{i}^{j})=
    - \mathbb{E}_{\hat{o_j}, a_{j}}\left[\log \pi_{i}^{j}\left(a_{j} | \hat{o_j}\right)
    + \lambda H(\pi_{i}^{j})\right]
} 
where $H(\pi_i^j)$ is the entropy of the policy distribution and $\lambda$ is a small weight (0.001 in experiments). The centralized critics $Q_i$ are updated with: 
\begin{align}
    \mathcal{L}\left(\theta_{i}\right) &=\mathbb{E}_{\mathbf{x}, a, r, \mathbf{x}^{\prime}}\left[\left(Q_{i}^{\mu}\left(\mathbf{x}, a_{1}, \ldots, a_{N}\right)-y\right)^{2}\right], \\ 
    y &=r_{i}+\gamma Q_{i}^{\mu}(\hat{o}^{\prime}, \mu_{i}^{\prime 1}\left(\hat{o_{1}}\right), \ldots, \mu_{i}^{\prime N} \left(\hat{o_{N}}\right))
\end{align}
where $\mu_i^{\prime}$ is a target policy and $\hat{o_i}^\prime$ is an inferred next observation following observation $o$.

\begin{wrapfigure}{R}{0.4\textwidth}
\centering
\vspace{-2em}
\includegraphics[width=.4\textwidth]{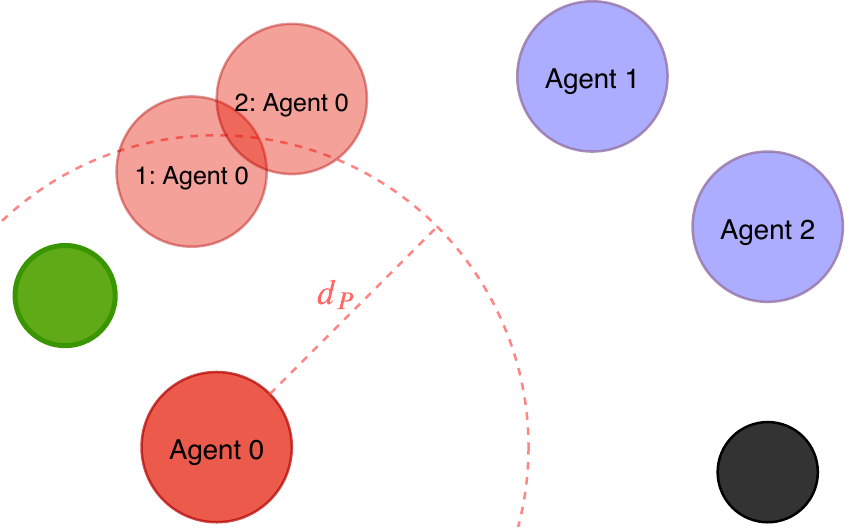}
\caption{Physical deception scenario with agent inference. 
Agents 1 and 2 are near and can observe each others' real positions. They are both too far from Agent 0 so they each sample the CC-WGAN to infer Agent 0's position.}
\label{fig:scenario_with_inference}
\vspace{-1em}
\end{wrapfigure}

\section{Experiments}

\subsection{Environments and Setup}
\label{sec:experiments}

We evaluate our method under three continuous scenarios of the Multi-agent Particles Environment (MPE)\footnote{MPE code: \url{http://github.com/openai/multiagent-particle-envs}} introduced in the original MADDPG paper \cite{Lowe2017}.
In order to directly compare to their results, we use the physical deception and predatory-prey competitive scenarios also used by \cite{Lowe2017}. We additionally test on a cooperative navigation scenario where agents share a reward function to show both competitive and cooperative settings. We did not include the communication-based scenarios tested by Lowe et al. because they have no clear way to make partially observable.
In contrast, we evaluate on physical scenarios that are easy to make partially observable: when agents' are farther from each other than partially observable distance $d_P$, they cannot observe each others' positions, velocity, etc. (see Fig. \ref{fig:scenario_with_inference}). We use $d_P=1$ in all experiments, where the width of the 2D square environment is 2. 

When using agent-distance partial observability, learning the coordination of predator-prey appears hardest, followed by physical deception, and lastly cooperative navigation. In predator-prey, three agents must coordinate to catch one faster agent, so there is no stable strategy. 
In physical deception, two agents should learn to deceive an adversary agent by covering two landmarks to hide which is the correct goal. 
If the adversary is out of range, this strategy should not change. 
In cooperative navigation, each agent must move to and remain near different landmarks. With myopic view $d_P$, agents can still determine if another agent is covering the same landmark and move to another. 

In addition to making the decentralized phase partially observable, we approximate real-world deployment by modifying simulation dynamics with scaled random normal noise and translation to both actions and observations. Combined with partial observability, the added noise makes learning more difficult for both the policies and the CC-WGAN and requires fine-tuning to the new distribution. 

Each episode has 200 steps with no early termination. All models are updated every 100 steps, and are represented with a three-layer, feed-forward neural network with 64 hidden units. The models use the Adam optimizer and each non-output layer uses a ReLU activation function. Each plot shows the mean and standard deviation shading over 30 independent trials for each algorithm. Each algorithm within a single plot receives the same set of 30 random seeds for accurate comparisons with random exploration. In all plots, a vertical dashed line marks the episode in which the environment becomes decentralized.

\begin{figure*}[ht]
\centering
\subfigure[Physical Deception]{\includegraphics[width=0.32\textwidth]{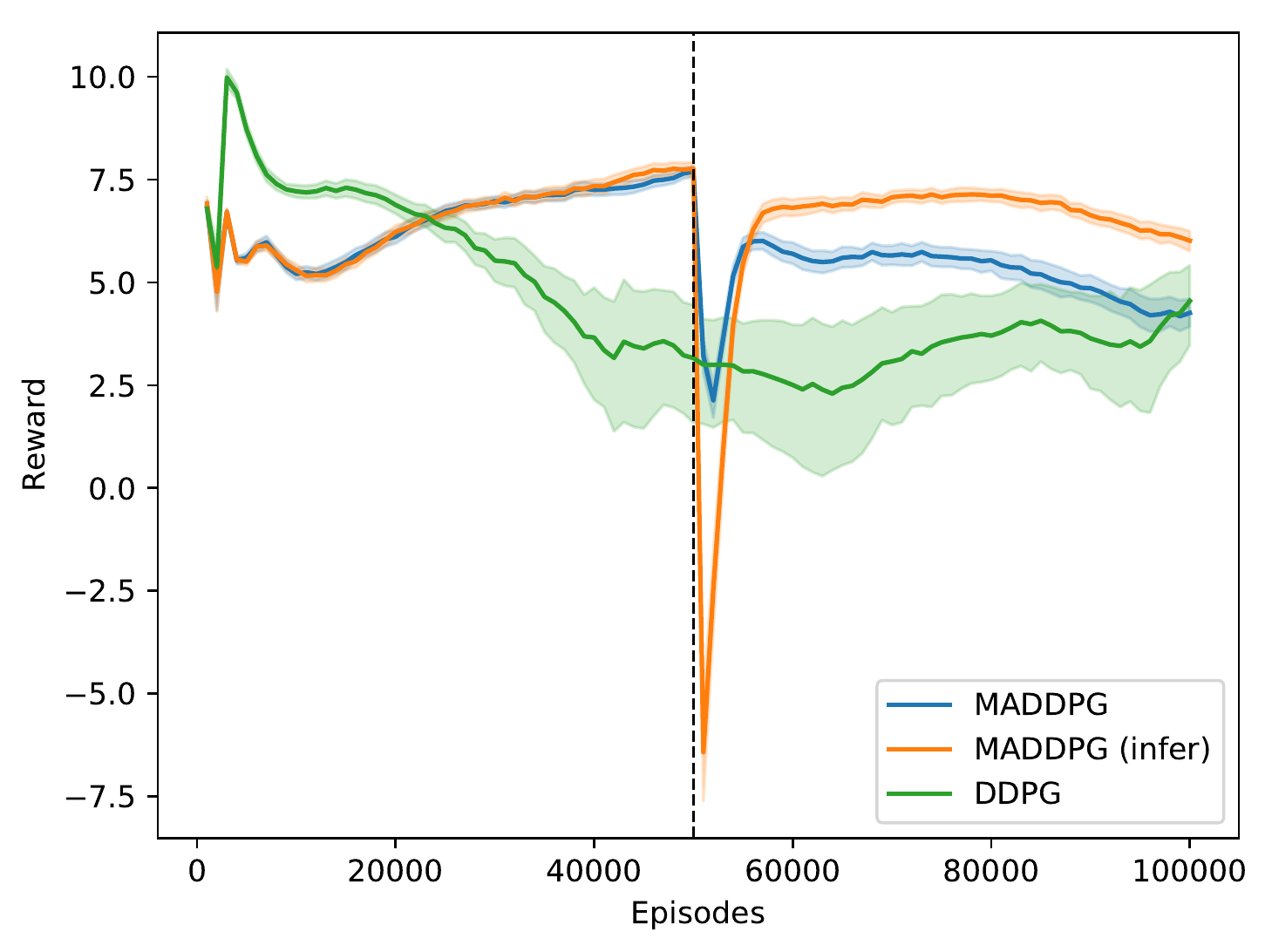}}
\subfigure[Predator-Prey]{\includegraphics[width=0.32\textwidth]{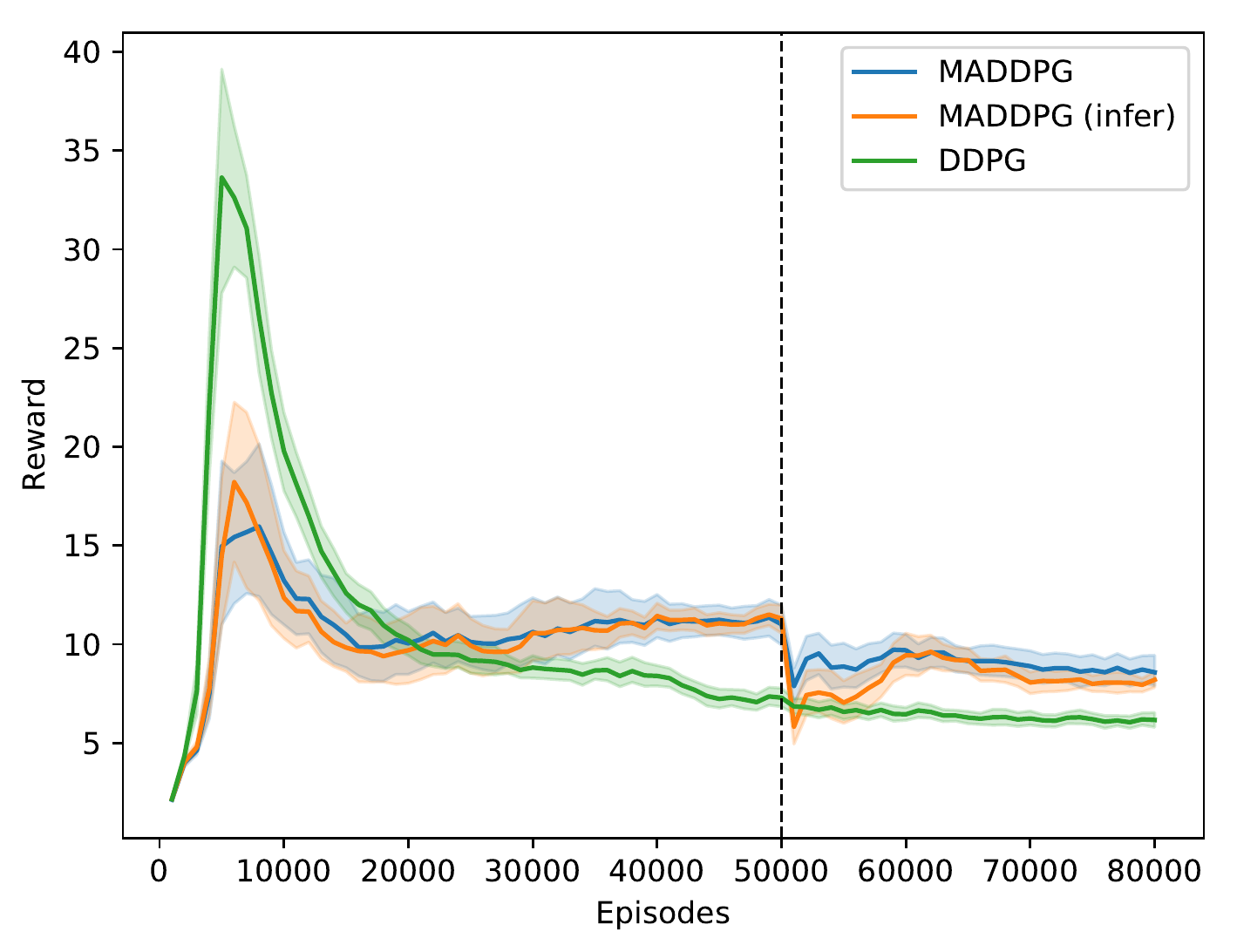}}
\subfigure[Cooperative Navigation]{\includegraphics[width=0.32\textwidth]{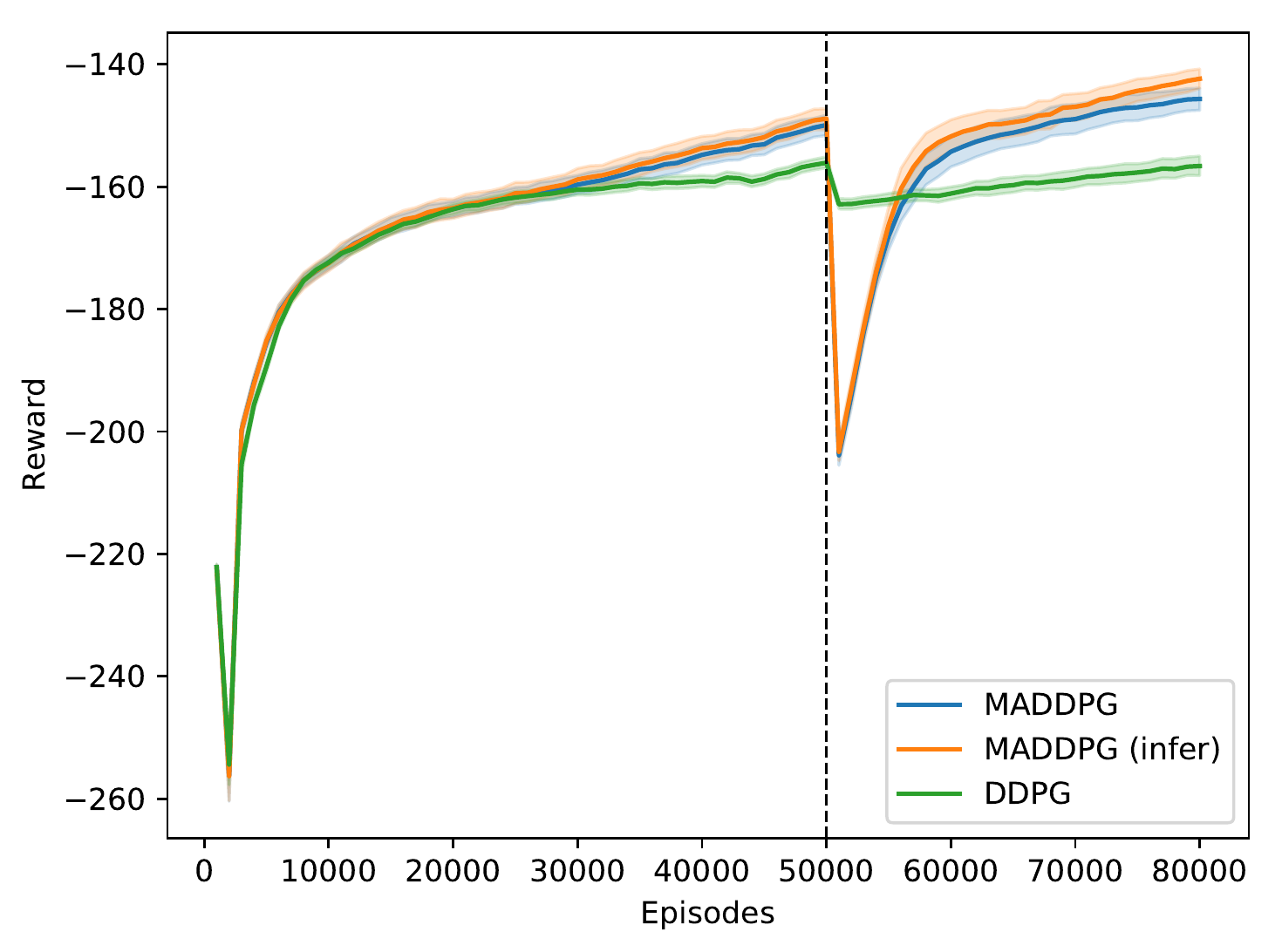}}
\caption{Reward for cooperating agents with distance-based partial observations in decentralized phase.}
\label{fig:PO_only}
\end{figure*}

\begin{figure*}[ht]
\centering
\subfigure[Physical Deception]{\includegraphics[width=0.32\textwidth]{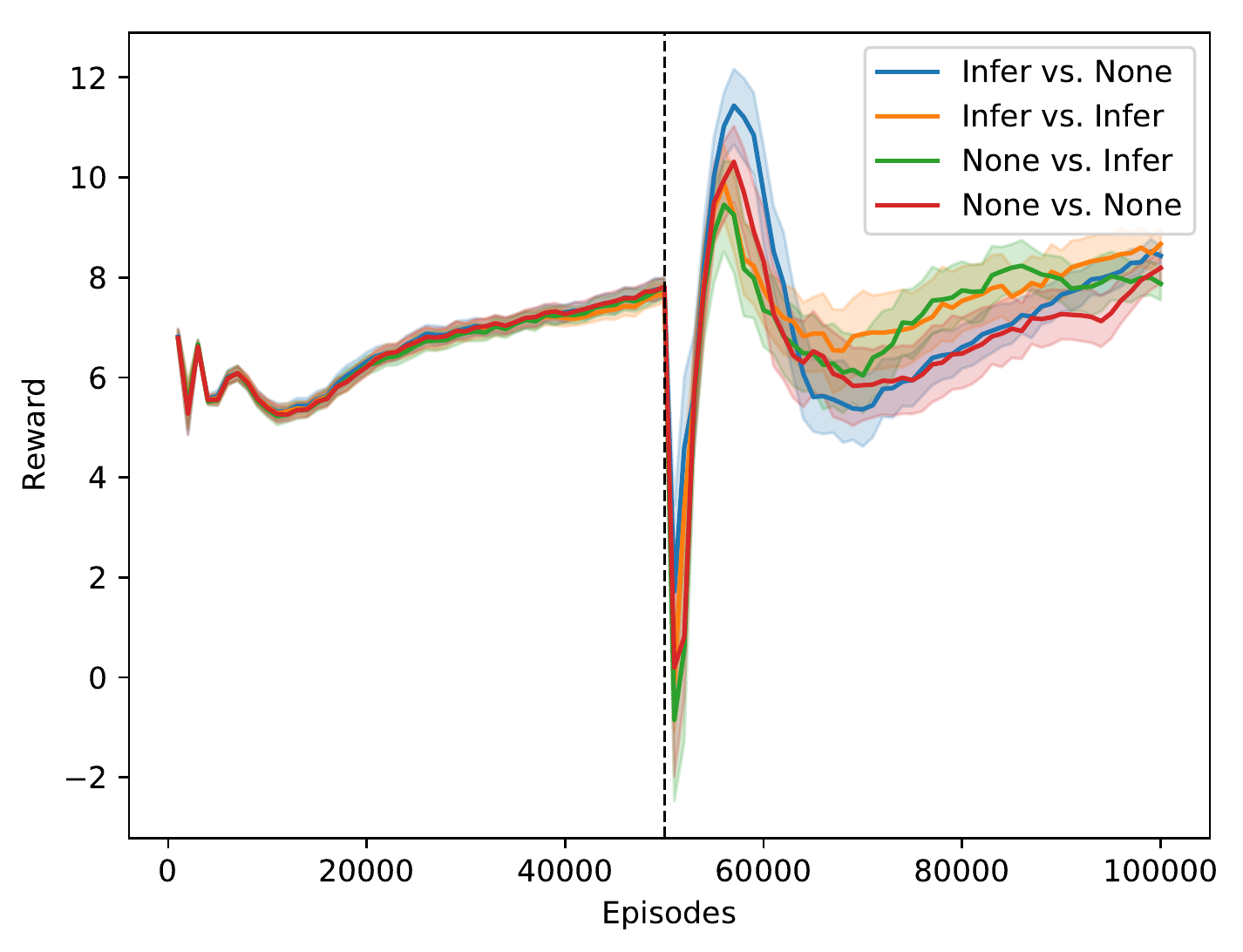}}
\subfigure[Predator-Prey]{\includegraphics[width=0.32\textwidth]{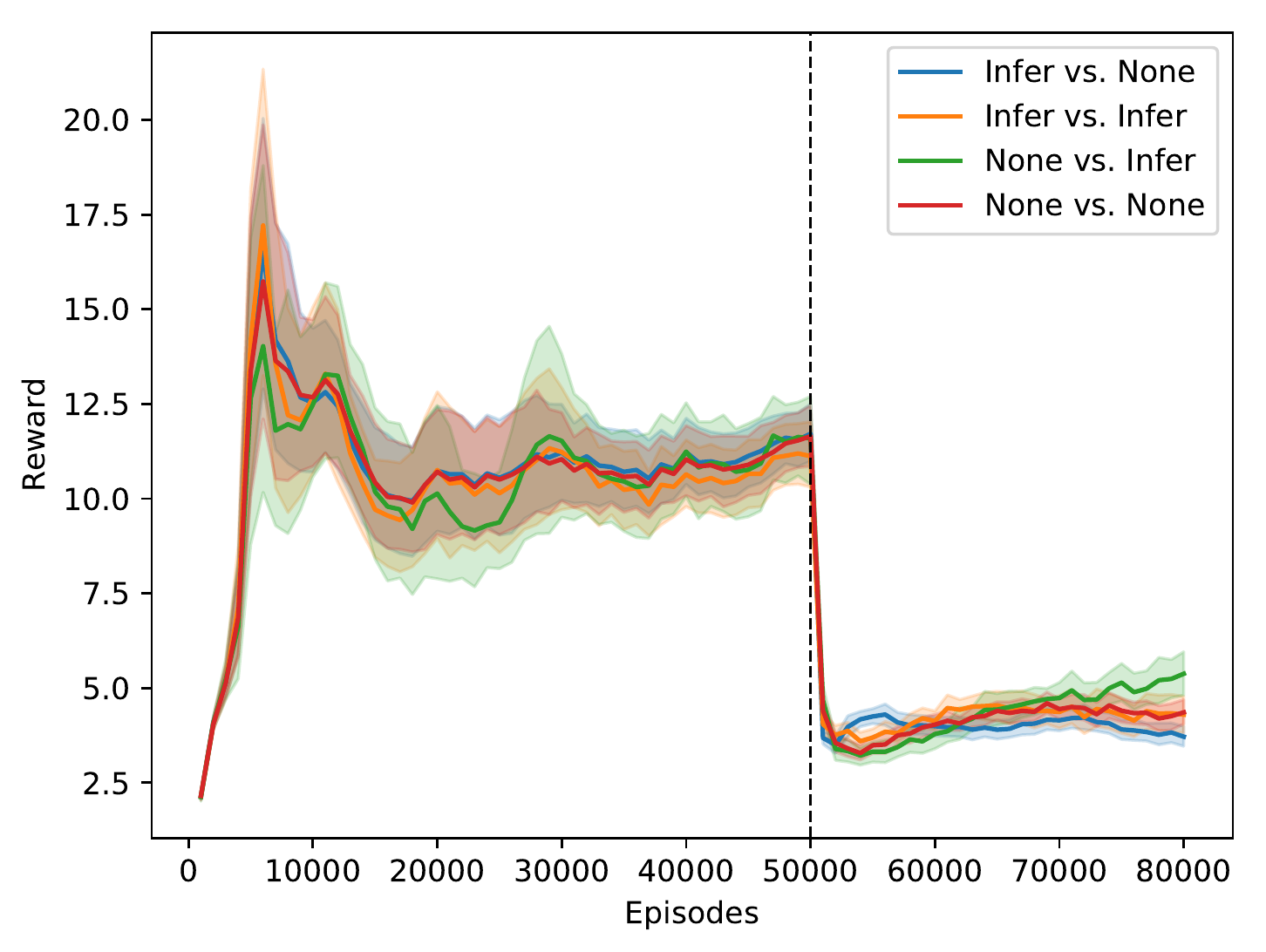}}
\subfigure[Cooperative Navigation]{\includegraphics[width=0.32\textwidth]{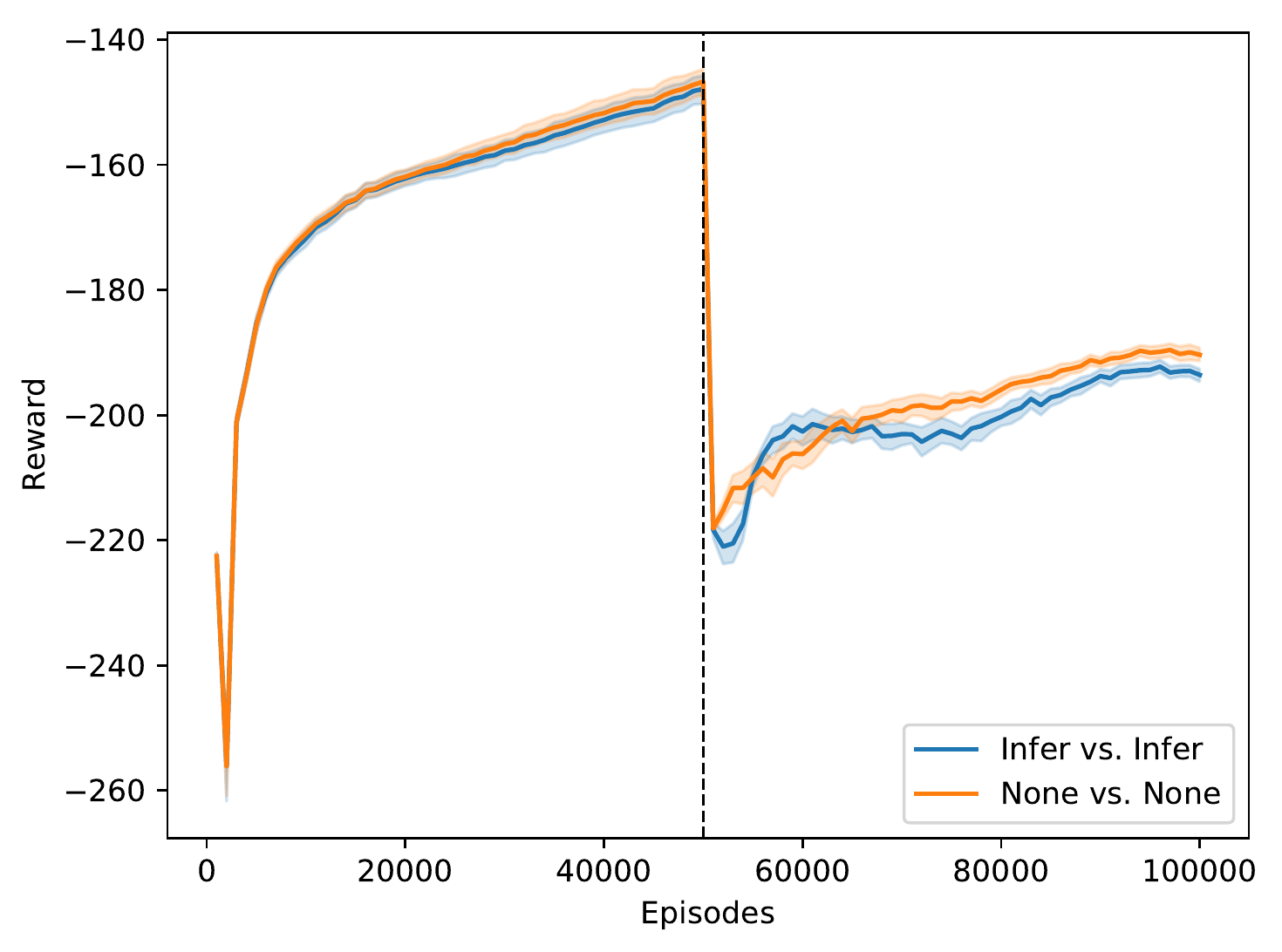}}
\caption{Both sides of cooperating agent rewards with both distance-based partial observations and altering environment dynamics in decentralized phase. ``Infer'' is our approach and ``None'' is MADDPG.}
\label{fig:PO_dynamics}
\end{figure*}

\begin{figure*}[ht]
\centering
\subfigure[Physical Deception]{\includegraphics[width=0.32\textwidth]{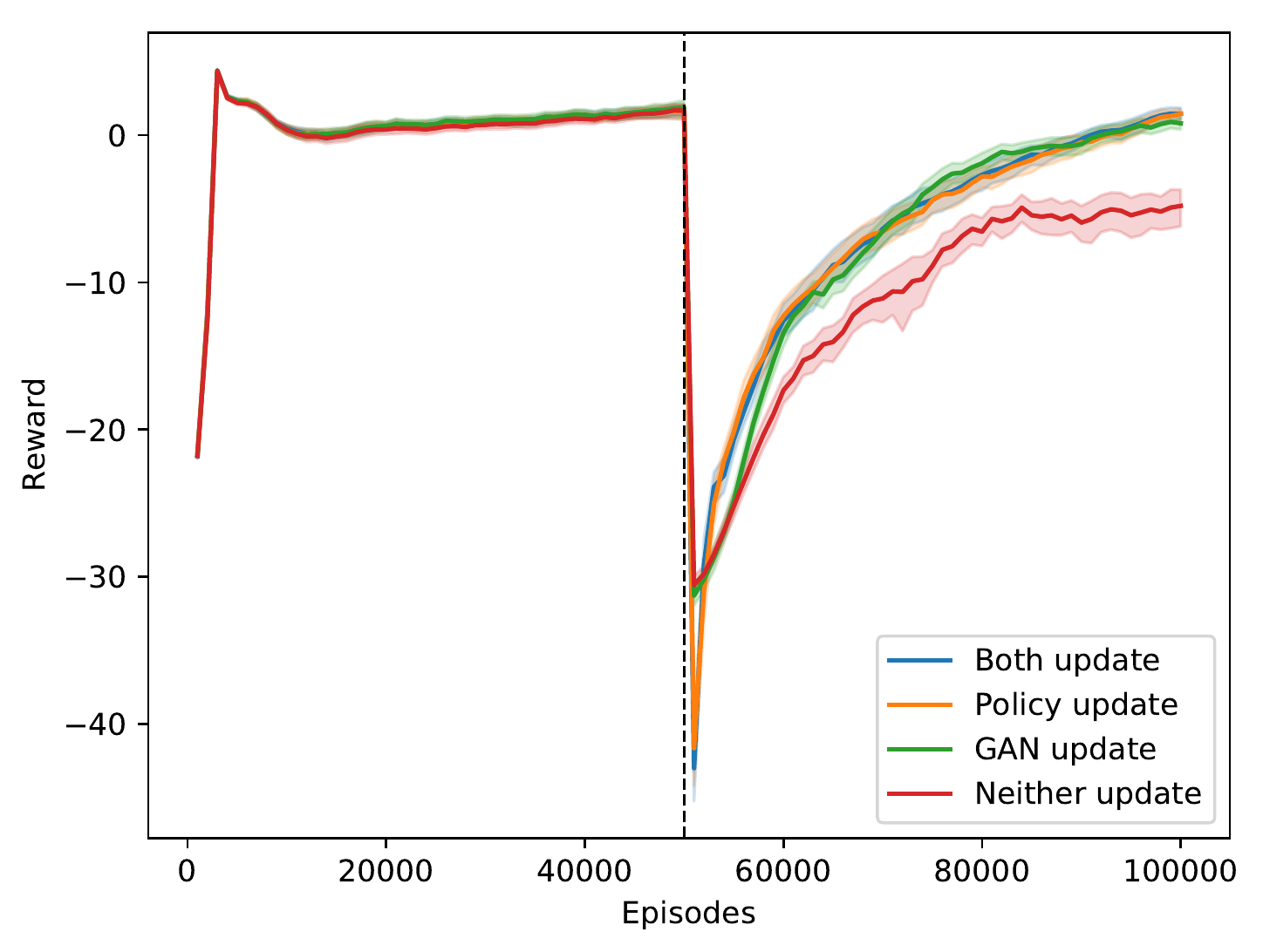}}
\subfigure[Predator-Prey]{\includegraphics[width=0.32\textwidth]{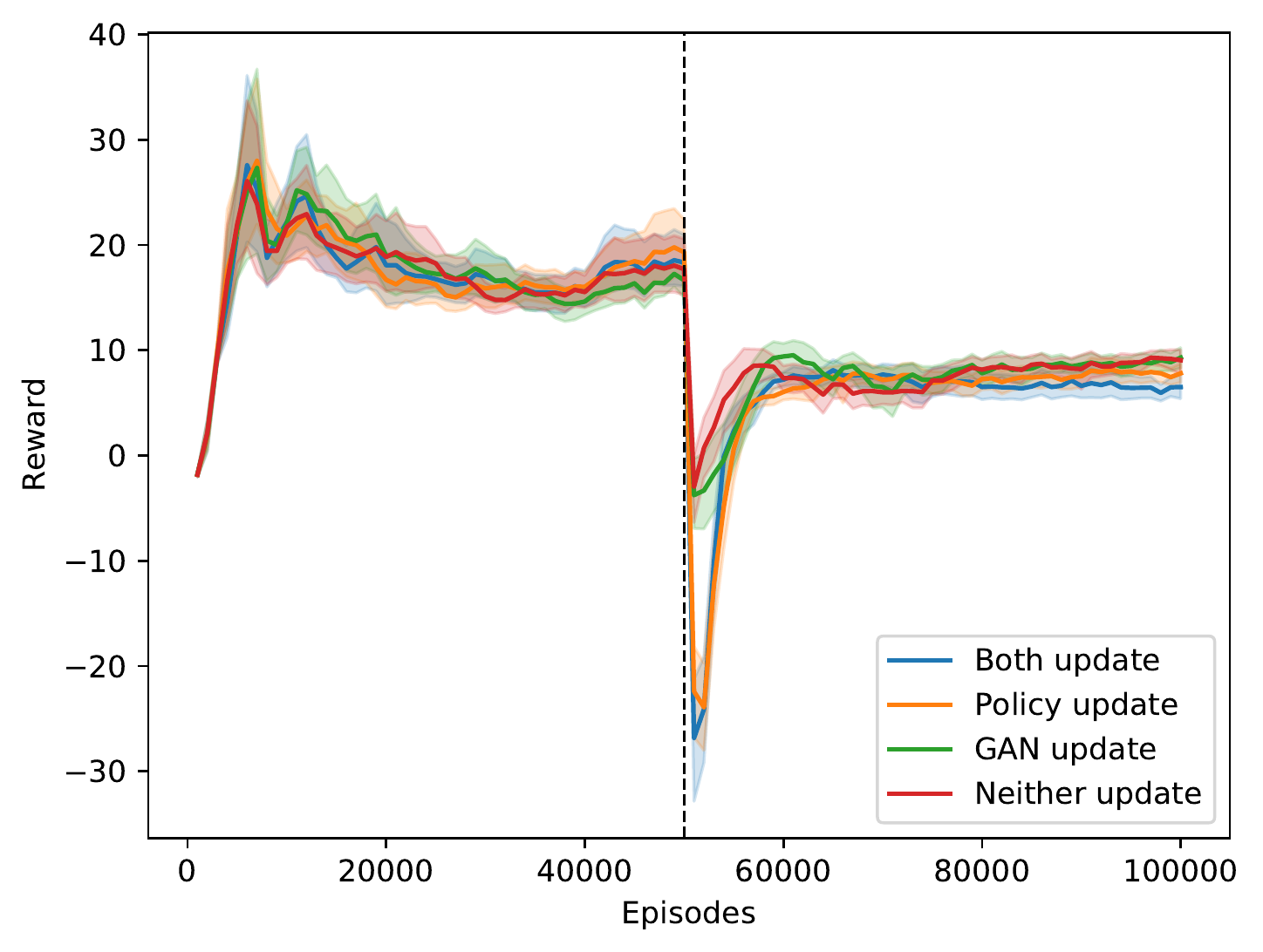}}
\subfigure[Cooperative Navigation]{\includegraphics[width=0.32\textwidth]{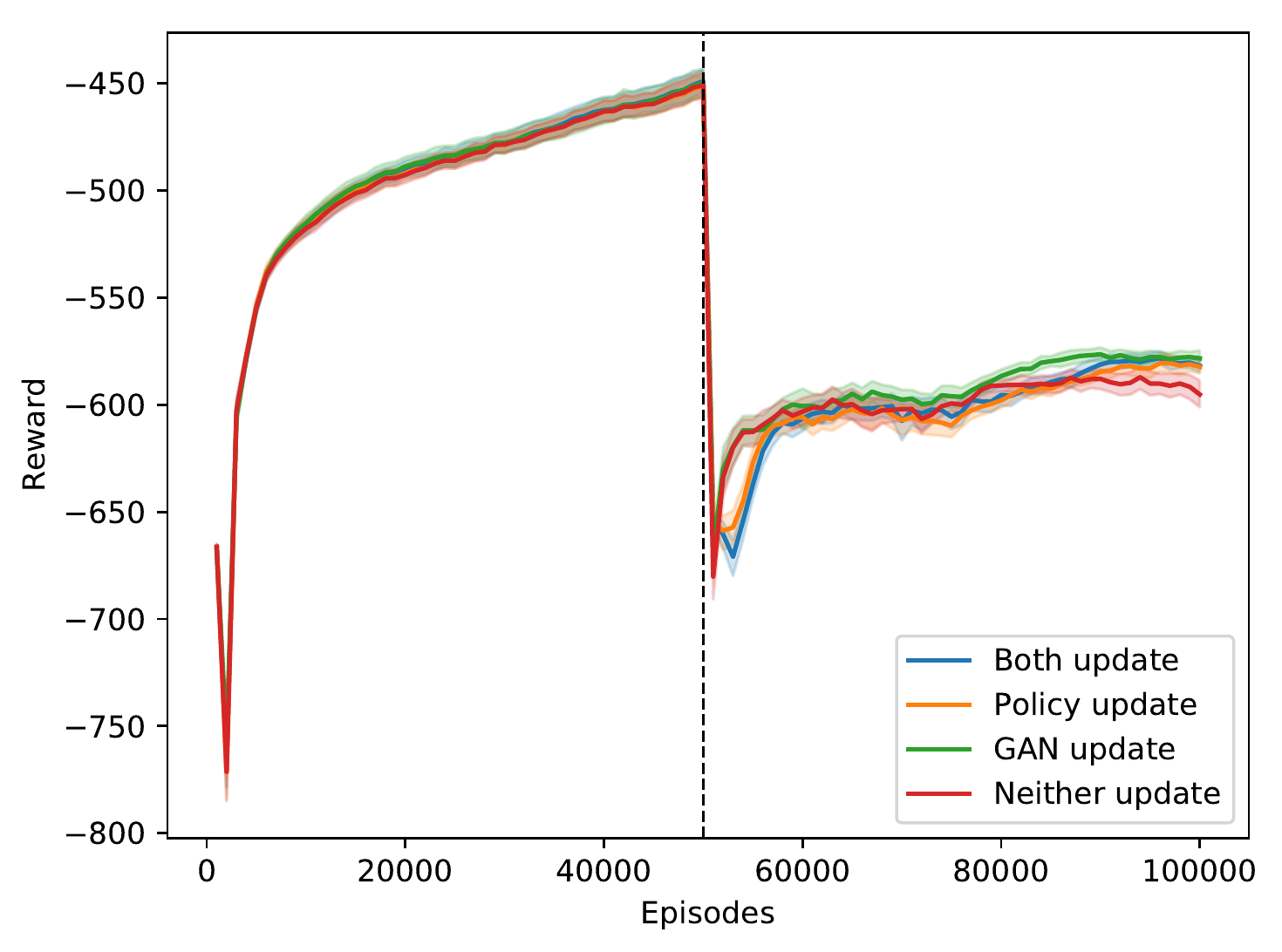}}
\caption{Total reward for our approach with all four combinations of whether agent policies and CC-WGAN continue to update on inferred observations during decentralized phase.}
\label{fig:dec_update}
\end{figure*}

\subsection{Results} 
\label{ssec:results}

The following plots compare our approach of augmenting MADDPG with CC-WGAN inference against regular MADDPG and DDPG. 
We chose DDPG because it performed the best among all IL methods in \cite{Lowe2017} and we use the same environment. 
Agents learn approximate policies for all other agents in MADDPG with and without generative inference. 
MADDPG and our version are identical during the centralized training because agents only use the CC-WGAN inference in the decentralized phase. 
After centralized training, we let MADDPG continue learning while treating the partial observability as random noise, whereas our approach infers the missing data. Except for Fig. \ref{fig:dec_update}, the policies and CC-WGAN continue updating in the decentralized phase. 

\begin{wrapfigure}{R}{0.38\textwidth}
\vspace{-1em}
\centering
\includegraphics[width=0.38\textwidth]{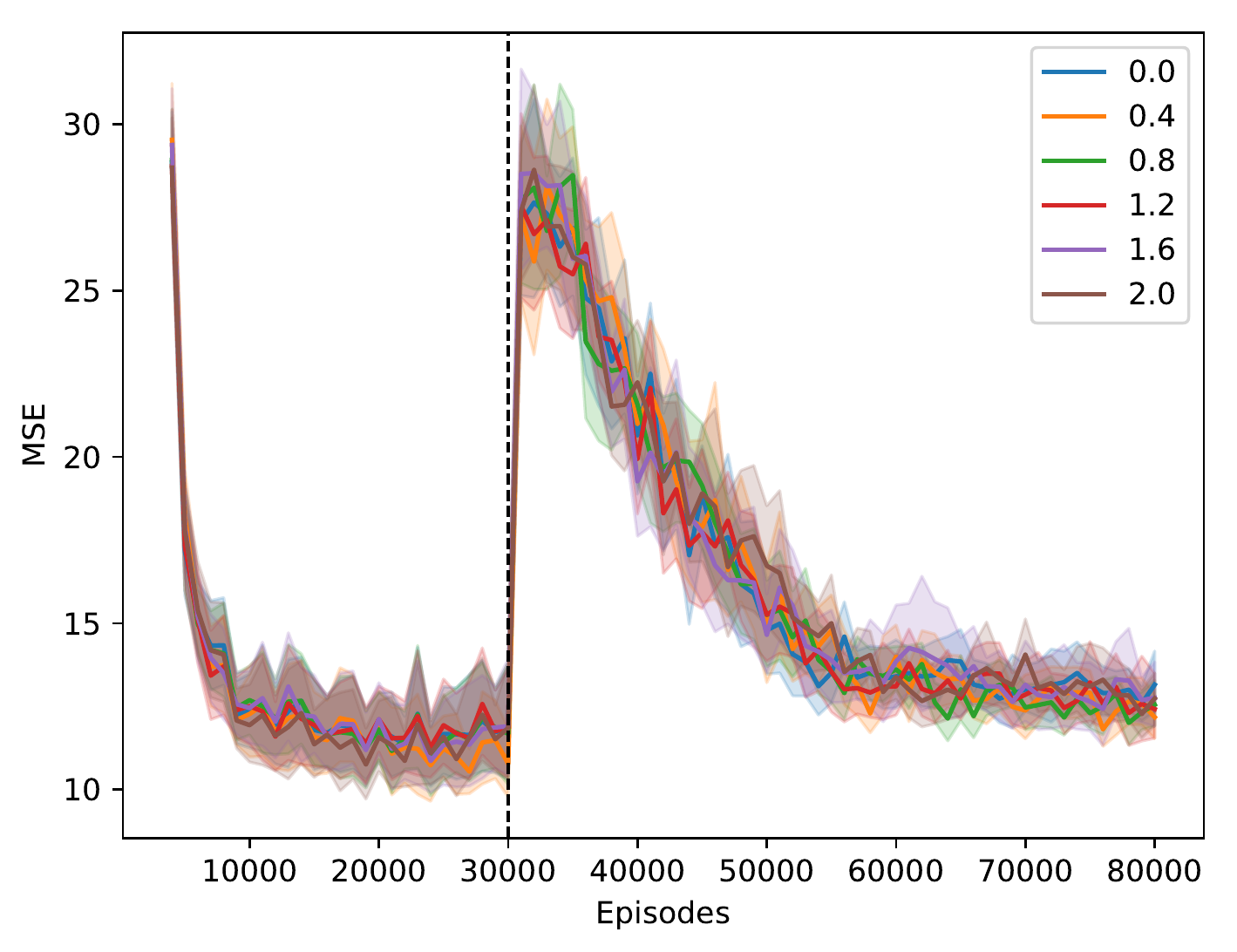}
\caption{CC-WGAN reconstruction MSE between latest joint observations $o$ and inferred observations $\hat{o}$ for predator-prey scenario. 
}
\label{fig:PO_dist_similarity}
\vspace{-1em}
\end{wrapfigure}

We did not test against the other CTDE methods discussed in Related Work because these methods use recurrent critics or policies. As such, they condition on past trajectories and would likely overcome the partial observability implicitly.


In Fig. \ref{fig:PO_only}, we show the cooperating agents' reward for all agents using MADDPG, MADDPG with CC-WGAN inference, and DDPG. In this plot we only introduce partial observability when agents are farther than the partially observable distance $d_P=1$. Fig. \ref{fig:PO_dynamics} shows the reward for cooperating agents with both partial observability and altered environment dynamics in the decentralized phase to evaluate the capability of fine-tuning to another environment. The overall reward is much lower here than Fig. \ref{fig:PO_only} which suggests the CC-WGAN is not well-suited to switching its modeled observation distribution (i.e., for sim-to-real transfer \cite{Peng2018}). 

In Fig. \ref{fig:dec_update}, we compare the total reward for our method with four options of decentralized updates, where the decentralized phase has both partial observability and altered environment dynamics. 
All agents use the CC-WGAN inferred observations when choosing actions. The curves show the difference in whether the policy and CC-WGAN update on inferred observations in the decentralized phase.

Lastly, in Fig. \ref{fig:PO_dist_similarity}, we show the CC-WGAN's reconstruction mean squared error during training over several partial observability distances $d_P \in [0.0, 2.0]$. When $d_P=0$, the model is effectively using IL in the decentralized phase; when $d_P=2$, the model is usually using complete observations. We show this reconstruction plot because the agents' observations are low-dimensional (i.e., not images as GANs are usually used for). 
MSE may not be a good metric for this error, however it was more informative than cosine similarity. 
We initially expected the error to be lower for larger $d_P$, but it is clear that the CC-WGAN reconstruction error has little to do with the partial observability distance. We would like to investigate the conditions under which the CC-WGAN gives better predictions.

\subsection{Discussion}
\label{ssec:discussion} 

The results shown here reveal properties about context-based modeling of observations in MARL and the scenarios for which our approach is appropriate. As CC-WGAN learns a joint observation distribution by sampling joint observations from its replay buffer \BG, it has no temporal coherence: each inference step is independent from the previous. Without a model of observation trajectories, it is ill-suited for dynamic tasks with no clear stable behaviors under the partial observability. 

Fig. \ref{fig:PO_only} illustrates this problem depending on the scenario's need for non-local information and the stability of optimal behavior. Inferring other agents' observations is useful when the task requires non-local coordination like the physical deception and predator-prey scenarios. 
Cooperative navigation agents can move to a different landmark if another agent is covering the same landmark, but may take slightly longer. Without temporal coherence, the CC-WGAN has trouble modeling non-stationary observation distributions like predator-prey. In contrast, physical deception and cooperative navigation have a stable optimal policy. In summation, our approach works best with a stable observation distribution (physical deception and cooperative navigation) and is useful in tasks requiring non-local coordination (physical deception and predator-prey). As such, our reward is significantly higher in physical deception, slightly higher in cooperative navigation, and slightly lower in predator-prey. 

As seen in Fig. \ref{fig:dec_update}, when the CC-WGAN and policy update on inferred observations the decentralized reward dropoff is more drastic than without updating on the inference. This is due to the co-adaptation between the agents' policies and CC-WGAN inference: the CC-WGAN must learn based on new observations being generated by agents choosing actions based on the inferred observations from the CC-WGAN. Also it appears that having either policy updates or GAN updates on inferred observations gives roughly the same benefit.

\section{Conclusions and Future Work}
\label{sec:conclusions}
In this paper we reviewed the recent trend of MARL methods utilizing centralized training with decentralized execution (CTDE) and identified that none of the methods may continue learning in the decentralize phase without adding explicit communication. We proposed to learn a context conditional generative model during centralized training phase that allows for a popular CTDE actor-critic method to continue learning in the decentralized phase, and showed that this addition allows for increased reward and coordination in three continuous multi-agent tasks. 

Our approach is useful for completing partial observations in Markovian environments where decentralized environment dynamics closely match the centralized training dynamics. In environments where agents should condition on history, recurrent policies or critics would help solve the problem. Our experiments show that context is useful in settings where there is a stable optimal behavior for agents, but training on trajectories may be able to learn more difficult observation distributions. 
We would also like to address the domain adaptation problem with possibly re-training the generative model on decentralized dynamics, or using techniques such as domain randomization.



\bibliographystyle{unsrt}
\bibliography{settings/references}

\begin{thebibliography}{10}

\bibitem{Mnih2013}
Volodymyr Mnih, David Silver, and Martin Riedmiller.
\newblock {Playing Atari with Deep Reinforcement Learning}.
\newblock {\em arXiv}, pages 1--9, 2013.

\bibitem{Haarnoja2018}
Tuomas Haarnoja, Sehoon Ha, Aurick Zhou, Jie Tan, George Tucker, and Sergey
  Levine.
\newblock {Learning to Walk via Deep Reinforcement Learning}.
\newblock In {\em International Conference on Machine Learning}, 2018.

\bibitem{Zhang2018a}
Kaiqing Zhang, Zhuoran Yang, Han Liu, Tong Zhang, and Tamer Basar.
\newblock {Fully Decentralized Multi-Agent Reinforcement Learning with
  Networked Agents}.
\newblock In {\em International Conference on Machine Learning}, 2018.

\bibitem{Denton2016}
Emily Denton, Sam Gross, and Rob Fergus.
\newblock {Semi-Supervised Learning with Context-Conditional Generative
  Adversarial Networks}.
\newblock {\em arXiv}, 2016.

\bibitem{Lowe2017}
Ryan Lowe, Yi~Wu, Aviv Tamar, Jean Harb, Pieter Abbeel, and Igor Mordatch.
\newblock {Multi-Agent Actor-Critic for Mixed Cooperative-Competitive
  Environments}.
\newblock In {\em Neural Information Processing Systems (NIPS)}, 2017.

\bibitem{Schmidhuber1997}
Jürgen Schmidhuber and Sepp Hochreiter.
\newblock {Long short-term memory}.
\newblock {\em Neural Computation}, 9(8):1735--1780, 1997.

\bibitem{Hausknecht2015}
Matthew Hausknecht and Peter Stone.
\newblock {Deep Recurrent Q-Learning for Partially Observable MDPs}.
\newblock In {\em 2015 AAAI Fall Symposium Series}, 2015.

\bibitem{Gupta2017}
Jayesh~K. Gupta, Maxim Egorov, and Mykel Kochenderfer.
\newblock {Cooperative Multi-agent Control Using Deep Reinforcement Learning}.
\newblock In {\em AAMAS Workshop}, volume 10642 LNAI, pages 66--83, 2017.

\bibitem{Foerster2018a}
Jakob~N Foerster, Gregory Farquhar, Triantafyllos Afouras, Nantas Nardelli, and
  Shimon Whiteson.
\newblock {Counterfactual Multi-Agent Policy Gradients}.
\newblock In {\em Association for the Advancement of Artificial Intelligence},
  2018.

\bibitem{Rashid2018}
Tabish Rashid, Mikayel Samvelyan, Christian Schroeder De~Witt, Gregory
  Farquhar, Jakob Foerster, and Shimon Whiteson.
\newblock {QMIX: Monotonic Value Function Factorisation for Deep Multi-Agent
  Reinforcement Learning}.
\newblock In {\em International Conference on Machine Learning}, 2018.

\bibitem{Foerster2016}
Jakob~N Foerster, Yannis~M Assael, Nando De~Freitas, and Shimon Whiteson.
\newblock {Learning to Communicate with Deep Multi-Agent Reinforcement
  Learning}.
\newblock In {\em Neural Information Processing Systems (NIPS)}, 2016.

\bibitem{Zhang2013}
Chongjie Zhang and Victor Lesser.
\newblock {Coordinating Multi-Agent Reinforcement Learning with Limited
  Communication}.
\newblock In {\em Autonomous Agents and Multi-Agent Systems}, 2013.

\bibitem{Zhang2018}
Wenxu Zhang, Lei Ma, and Xiaonan Li.
\newblock {Multi-agent reinforcement learning based on local communication}.
\newblock {\em Cluster Computing}, 2018.

\bibitem{Littman1994}
Michael~L Littman.
\newblock {Markov games as a framework for multi-agent reinforcement learning}.
\newblock In {\em Machine Learning}, 1994.

\bibitem{sutton2018reinforcementbook}
Richard~S Sutton and Andrew~G Barto.
\newblock {\em {Reinforcement Learning: An Introduction}}.
\newblock MIT press, 2nd edition, 2018.

\bibitem{Silver2014}
David Silver, Nicolas Heess, Thomas Degris, Daan Wierstra, and Martin
  Riedmiller.
\newblock {Deterministic Policy Gradient Algorithms}.
\newblock In {\em International Conference on Machine Learning}, 2014.

\bibitem{Lillicrap2016a}
Timothy~P Lillicrap, Jonathan~J Hunt, Alexander Pritzel, Nicolas Heess, Tom
  Erez, Yuval Tassa, David Silver, and Daan Wierstra.
\newblock {Continuous control with deep reinforcement learning}.
\newblock In {\em International Conference on Learning Representations}, 2016.

\bibitem{Goodfellow2014}
Ij~Goodfellow, J~Pouget-Abadie, and Mehdi Mirza.
\newblock {Generative Adversarial Networks}.
\newblock {\em arXiv preprint arXiv: {\ldots}}, pages 1--9, 2014.

\bibitem{Arjovsky2017}
Martin Arjovsky, Soumith Chintala, and Léon Bottou.
\newblock {Wasserstein GAN}.
\newblock {\em arXiv}, 2017.

\bibitem{VanOord2016}
Aaron van Oord, Nal Kalchbrenner, and Koray Kavukcuoglu.
\newblock {Pixel Recurrent Neural Networks}.
\newblock In {\em International Conference on Machine Learning}, pages
  1747--1756, 2016.

\bibitem{Pathak2016}
Deepak Pathak, Philipp Kr{\"{a}}henb{\"{u}}hl, Jeff Donahue, Trevor Darrell,
  and Alexei~A Efros.
\newblock {Context Encoders: Feature Learning by Inpainting}.
\newblock {\em arXiv}, 2016.

\bibitem{Gulrajani2017}
Ishaan Gulrajani, Faruk Ahmed, Martin Arjovsky, Vincent Dumoulin, and Aaron
  Courville.
\newblock {Improved Training of Wasserstein GANs}.
\newblock In {\em Advances in Neural Information Processing Systems}, pages
  5767--5777, 2017.

\bibitem{Goodfellow2016book}
Ian Goodfellow, Yoshua Bengio, and Aaron Courville.
\newblock {\em {Deep Learning}}.
\newblock MIT press, 2016.

\bibitem{Peng2018}
Xue~Bin Peng, Marcin Andrychowicz, Wojciech Zaremba, and Pieter Abbeel.
\newblock {Sim-to-Real Transfer of Robotic Control with Dynamics
  Randomization}.
\newblock In {\em International Conference on Robotics and Automation}, 2018.

\end{thebibliography}

\end{document}